\documentstyle{article}
\title{The Atiyah-Singer fixed point theorem and the gauge field copy problem\thanks{This paper was published in Bol. Soc. Paran. Matem. vol. 16 59-62 (1996).}}
\author{Adonai S. Sant'Anna\\{\small Department of Mathematics, Federal University at Paran\'a}\\{\small P.O. Box 19.081, Curitiba, PR, 81530-900 Brazil.}}
\date{ }
\begin{document}
\maketitle
\newcounter{cms}
\setlength{\unitlength}{1mm}
\begin{quote}
{\bf Abstract} - {\small A condition for the existence of false gauge field copies in terms of the Lefschetz number of a certain differential operator is presented.}
\end{quote}

\begin{quote}
\end{quote}

\section{Introduction}
\newtheorem{definicao}{Definition}[section]
\newtheorem{teorema}{Theorem}[section]
\newtheorem{lema}{Lemma}[section]
\newtheorem{corolario}{Corolary}[section]
\newtheorem{proposicao}{Proposition}[section]
\newtheorem{axioma}{Axiom}[section]
\newtheorem{observacao}{Observation}[section]

	Gauge field theories may be defined as physical interpretations of the theory of connections in a principal fiber bundle \cite{Cho}. Some gauge fields (curvature forms) admit two or more potentials (connection forms), which are related to the fields by the equation field (structure equation). Such an ambiguity is known as the gauge field copy problem and it was discovered in 1975 by T.T. Wu and C.N. Yang \cite{Wu}. Gauge copies fall into two cases: `true copies' (potentials that are not locally related by a gauge transformation) and `false copies' (locally gauge equivalent potentials).

	Our goal, in this paper, is to establish an analytical condition for the existence of false gauge field copies. We use a result due to F.A. Doria \cite{Doria-84}, where a topological condition to the existence of false gauge field copies is presented. The Atiyah-Singer Fixed Point Theorem \cite{Shanahan} helps to make the connection between the topological condition and the analytical one.

	We use standard notation. For details see \cite{Bleecker}, \cite{Kobayashi}, and \cite{Shanahan}.

	Let $P(M,G)$ be a principal fiber bundle, where $M$ is a finite-dimensional smooth real manifold and $G$ is a finite-dimensional Lie group. If we denote by $(P,\alpha)$ the principal fiber bundle $P(M,G)$ endowed with the connection-form $\alpha$, and by $L$ the field that corresponds to the potential $A$ associated to $\alpha$, then:

\begin{definicao}
The field $L$ or the potential $A$ are reducible if the corresponding bundle $(P,\alpha)$ is reducible.\label{def-redutibilidade}
\end{definicao}

\begin{teorema}
Let $P(M,G)$ be as previously stated but with the extra condition that $G$ is semi-simple. $L$ has potentials that are locally related by a gauge transformation if and only if $L$ is reducible.\label{Teo-Doria}
\end{teorema}

{\bf Proof}: See \cite{Doria-84}.$\Box$

\section{The Atiyah-Singer Fixed Point Theorem}

	Let $G$ be a compact Lie group acting on a smooth manifold $X$, and let $D$ be a $G$-invariant elliptic partial differential operator on $X$. We can now state the Atiyah-Singer fixed point theorem \cite{Shanahan}:

\begin{teorema}
The Lefschetz number $L(g,D)$ is related to the fixed point set $X^{g} = \{x\in X; gx = x\}$ by the formula

\begin{equation}
L(g,D) =
(-1)^{m}\left(\frac{ch_{g}(j^{*}\sigma(D))}{ch_{g}
(\bigwedge_{-1}^{g}N^{g}\otimes {\bf C})}td(T^{g}\otimes{\bf
C})\right) [TX^{g}],
\end{equation}
where $m$ is the dimension of $X^{g}$, $j^{*}:K_{G}(TX)\rightarrow K_{G}(TX^{g})$ is induced by the inclusion mapping $j:X^{g}\rightarrow X$, $\sigma(D)$ is the symbol of $D$ (and so it is an element of the Grothendieck Group $K_{G}(TX)$), $N^{g}$ is the normal bundle $NX^{g}$ of $X^{g}$ in $X$, $T^{g}$ is the tangent bundle $TX^{g}$ of $X^{g}$ in $X$, $ch_{g}$ is the Chern Character, $td$ is the Todd class, $\bigwedge_{-1}^{g}$ is the Thom class, and {\bf C} denotes the topological field of complex numbers.\label{Lefschetz}
\end{teorema}

\section{An Analytical Condition For The Existence of False Gauge Field Copies}

	Gauge fields and gauge potentials can be defined as cross-sections of vector bundles associated with the principal fiber bundle $P(M,G)$. The potential space (or connection space) coincides with the space of all $C^{k}$ cross-sections of the vector bundle $E$ of $l(G)$-valued 1-forms on $M$, where $l(G)$ is the group's Lie algebra, while the field space (or curvature space) coincides with the space of all $C^{k}$ cross-sections of the vector bundle ${\bf E}$ of $l(G)$-valued 2-forms on $M$.

	Let $F$ and ${\bf F}$ be manifolds on which $G$ acts on the left and such that $E = P\times_{G}F$ and ${\bf E} = P\times_{G}{\bf F}$, where $P$ is the total space of $P(M,G)$. In other words, $E$ is the quotient space of $P\times F$ by the group action. Similarly, ${\bf E}$ is the quotient space of $P\times {\bf F}$ by the action of the group G.

	Before the statement of our main result, we must recall that a topological group $G$ acts freely on a topological space $X$ if and only if $xg = x$ implies that $g = e$, where $x\in X$, $g\in G$, and $e$ is the identity element in $G$.

\begin{teorema}
If ${\cal D}_{G}:C^{\infty}(P;P\times F)\rightarrow
C^{\infty}(P;P\times {\bf F})$ is a $G$-invariant elliptic partial differential operator, then the Lefschetz number $L(g,{\cal D}_{G})$ can be defined if and only if $g=e$.\label{1}
\end{teorema}

{\bf Proof:} As $G$ acts freely on $P$, then $P^{e} = P$
and $P^{g} = \emptyset$ if $g\neq e$. Hence, $j^{*}\sigma({\cal D})$ (which is necessary to compute the Lefschetz number of ${\cal D}$) is not defined for $g\neq e$, since $TP^{g}=\emptyset$.$\Box$\\

	Let's abbreviate $L(e,{\cal D}_{G})$ as $L({\cal D}_{G})$. With theorem \ref{1} in mind, we establish the following result:

\begin{teorema}
If a gauge field (a cross-section of ${\bf E}$) is associated to copied potentials that are locally gauge-equivalent, then there is: a non-trivial sub-group of $G$, denoted by $G'$; a $G'$-manifold $P'$; and two $G'$-vector spaces $F'$ and $\bf F'$ such that if there is an elliptic partial differential operator\\
\begin{equation}
{\cal D}_{G'}:C^{\infty}(P';P'\times F')\rightarrow
C^{\infty}(P';P'\times {\bf F'})\label{8}
\end{equation}
$G'$-invariant, then its Lefschetz number $L({\cal D}_{G'})$ can be defined as a function of $G$-spaces.\label{II}
\end{teorema}

{\bf Proof:}  If a gauge field is associated to copied potentials that are locally gauge-equivalent, then such a field is reducible (Theorem \ref{Teo-Doria}). Therefore $P(M,G)$ is reducible (Definition \ref{def-redutibilidade}). So, there is a non-trivial sub-group $G'$ of $G$ and a monomorphism $\varphi:G'\rightarrow G$ such that it can
be defined a reduced principal fiber bundle $P'(M,G')$ and a
reduction $f:P'(M,G')\rightarrow P(M,G)$. So, there are induced reductions f: $TP'\rightarrow TP$, and {\bf f}: $NP'\rightarrow NP$. If we consider $\varphi^{*}:K_{G}(TP)\rightarrow K_{G'}(TP')$, $\phi^{*}:K_{G}(N\otimes{\bf C})\rightarrow K_{G'}(N'\otimes {\bf C})$, and $\Phi^{*}:K_{G}(T\otimes{\bf C})\rightarrow K_{G'}(T\otimes {\bf C})$ as monomorphisms induced by $\varphi$, then, in accordance with theorem \ref{Lefschetz}:

\begin{equation}
L({\cal D}_{G'}) =
(-1)^{m}\left(\frac{ch_{e}(j^{*}\varphi^{*}\sigma({\cal D}_{G}))}{ch_{e}
(\bigwedge_{-1}\phi^{*} N\otimes {\bf C})}td(\Phi^{*} T\otimes{\bf C})\right)\varphi^{*}[TP].\label{*}
\end{equation}
$\Box$\\

	Theorem \ref{Teo-Doria} is a topological condition for false gauge field copies since it deals with the concept of reducibility to sub-groups of a topological group. Theorem \ref{II} refers to the Lefschetz number of an elliptic partial differential operator, which characterizes an analytical condition. An obvious corollary may be obtained with respect to the $G$-signature theorem \cite{Shanahan}. But this is left as an exercise to the reader.

	Other theorems \cite{SDC} \cite{Sant'Anna} may be obtained by the use, e.g., of the Atiyah-Singer index theorem \cite{Atiyah-Index-I} \cite{Shanahan}, in the sense of obtaining an analytical condition for the existence of gauge field copies. We are working also in a generalization of such ideas for the case of true gauge field copies.  If we modify the geometry of an irreducible principal fiber bundle in a manner to handle true copies as false \cite{Costa-Amaral}, it seems possible to state topological and analytical conditions for generic gauge field copies.

\section{Aknowledgements}

	This paper was partially prepared during a stay at Stanford University as a post-doctoral fellow. I would like to thank Patrick Suppes for his kind hospitality.

	I thank also the financial support from CNPq.

\end{document}